\documentclass[journal=jacsat,manuscript=article]{achemso}
\setkeys{acs}{articletitle = true}
\usepackage[version=3]{mhchem}
\usepackage[T1]{fontenc}
\usepackage{amsmath}
\usepackage{dcolumn}
\usepackage{amssymb,amsmath}
\usepackage{color}
\usepackage{epsfig}
\usepackage{graphicx}
\usepackage{epstopdf}

\author{Shang-Hsuan Wu}
\affiliation[Academia Sinica]
{Research Center for Applied Sciences, Academia Sinica, Taipei 11529, Taiwan}
\author{Ming-Yi Lin}
\affiliation[Chung Yuan Christian University]
{Department of Electronic Engineering, Chung Yuan Christian University, Taoyuan 32023, Taiwan}
\author{Sheng-Hao Chang}
\affiliation[Chung Yuan Christian University]
{Department of Electronic Engineering, Chung Yuan Christian University, Taoyuan 32023, Taiwan}
\author{Wei-Chen Tu}
\affiliation[Chung Yuan Christian University]
{Department of Electronic Engineering, Chung Yuan Christian University, Taoyuan 32023, Taiwan}
\author{Chih-Wei Chu}
\affiliation[Academia Sinica]
{Research Center for Applied Sciences, Academia Sinica, Taipei 11529, Taiwan}
\alsoaffiliation[Chang Gung University]
{College of Engineering, Chang Gung University, Taoyuan 33302, Taiwan}
\email{gchu@gate.sinica.edu.tw}
\author{Yia-Chung Chang}
\affiliation[Academia Sinica]
{Research Center for Applied Sciences, Academia Sinica, Taipei 11529, Taiwan}
\email{yiachang@gate.sinica.edu.tw}

\title[An \textsf{achemso} demo]
{A Design Based on Stair-case Band Alignment of Electron Transport Layer for Improving Performance and Stability in Planar Perovskite Solar Cells\footnote{A Design Based on Stair-case Band Alignment of Electron Transport Layer for Improving Performance and Stability in Planar Perovskite Solar Cells}}

\abbreviations{IR,NMR,UV}

\begin{document}

\begin{tocentry}

\centering
\includegraphics[width=6.5cm]{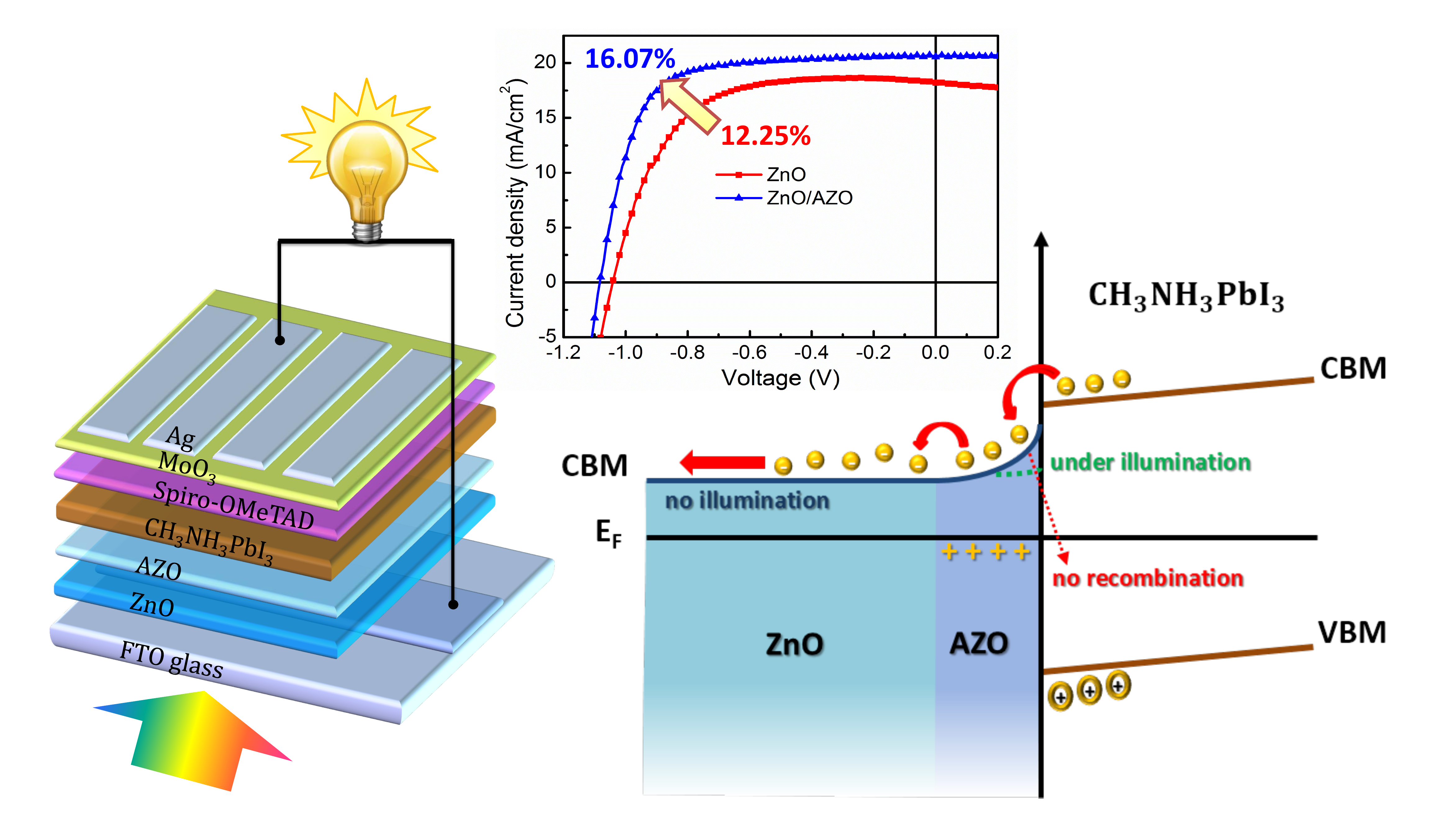}

\end{tocentry}

\begin{abstract}

Among the n-type metal oxide materials used in the planar perovskite solar cells, zinc oxide (ZnO) is a promising candidate to replace titanium dioxide (TiO$_2$) due to its relatively high electron mobility, high transparency, and versatile nanostructures. Here, we present the application of low temperature solution processed ZnO/Al-doped ZnO (AZO) bilayer thin film as electron transport layers (ETLs) in the inverted perovskite solar cells{, which provide a stair-case band profile.} Experimental results revealed that the power conversion efficiency (PCE) of perovskite solar cells were significantly increased from 12.25 to 16.07\% by employing the AZO thin film as the buffer layer. {Meanwhile, the short-circuit current density ($J_{sc}$), open-circuit voltage ($V_{oc}$), and fill factor (FF)  were improved  to 20.58 mA/cm$^2$, 1.09V, and 71.6\%, respectively.} The enhancement in performance { is attributed to the modified interface in ETL with stair-case band alignment of ZnO/AZO/CH$_3$NH$_3$PbI$_3$, which allows  more efficient extraction of photogenerated electrons in the} CH$_3$NH$_3$PbI$_3$ active layer. Thus, it is demonstrated that the ZnO/AZO bilayer ETLs would benefit the electron extraction and contribute in enhancing the performance of perovskite solar cells.

\end{abstract}

\section{Introduction}

Organometal halide perovskite solar cells (e.g., CH$_3$NH$_3$PbX$_3$, X = Cl, Br, I) have drawn much attention in current renewable solar research owing to their excellent optical and electronic properties{, including} strong absorption band that span the visible region\cite{Wolf2014}, direct band gap ($\sim1.5$ eV)\cite{Kim2012}, long carrier diffusion length (100 -- 1000 nm)\cite{Stranks2013}, high charge-carrier mobility ($\sim$10 cm$^2$V$^{-1}$s$^{-1}$),\cite{Wehrenfennig2014} and significantly low-cost fabrication process\cite{You2014,Wojciechowski2013}. These superior optoelectronic properties enable an increase in power conversion efficiencies (PCEs) of planar heterojunction (PHJ) perovskite devices from 3.8\% in 2009\cite{Kojima2009} to 22.1\% in 2016\cite{NREL2016}. {Due to} the potential of organometal halide perovskites, there have been a great number of studies {regarding the carrier} dynamics of perovskites\cite{Zhao2013}, development of new materials\cite{Lee2012,Yang2015} , optimization of perovskite absorber layer thickness, crystallinity and surface coverage \cite{Burschka2013,Eperon2013,Liu2013}, and low-temperature {processing}\cite{Ball2013,Wang2014} over the past few years. The state-of-art perovskite solar cells are based on two different device architectures: mesoporous\cite{Etgar2012} and planar heterojunctions \cite{Chen2013,Kang2014,Moorthy2016}. For the n-type metal oxide materials in the cell, titanium dioxide (TiO$_2$) is commonly employed in the mesoporous scaffold. Although TiO$_2$-based perovskite solar cells may pave the way to high PCE devices, several disadvantages of TiO$_2$ were reported such as low electron mobility and high annealing temperature (above 450$^{\circ}$C) in order to {form crystalline TiO$_2$ film in} anatase phase\cite{Beiley2012}. Accordingly, among other n-type metal oxide materials used in inverted solar cells, zinc oxide (ZnO) has been extensively studied as {a} substitute to TiO$_2$ due to its similar {electron affinity}, relatively high electron mobility, high transparency, and versatile nanostructures\cite{Ozgur2005,Wu2016}.

For high performance inverted perovskite solar cell, the selection of the electron collection layer {with} hole blocking capability and low resistivity pathway for efficient electron extraction is necessary. ZnO have been demonstrated to be an effective electron collection material due to its various nanostructures that can be easily achieved by solution process for more efficient charge extraction and transport\cite{Sekine2009,Ambade2015}. Recently, several studies of mesostructured perovskite solar cell based on ZnO nanorod arrays have been used to replace the mesoporous TiO$_2$ nanostructures in the conventional perovskite solar cells\cite{Dong2014,Liang2014}. For instance, Mahmood et al. have obtained remarkably PCE of 16.1\% in ZnO mesoscopic perovskite solar cells based on synergistically combining mesoscale control with nitrogen doping and surface modification of ZnO nanorod arrays\cite{Mahmood2015}. However, it was found that severe decomposition of CH$_3$NH$_3$PbI$_3$ could occur at the {ZnO}/perovskite interface due to {excessive} OH$^-$ groups and chemical residuals on the ZnO surface\cite{Cheng2015}. To overcome the unstable decomposition process of CH$_3$NH$_3$PbI$_3$ to PbI$_2$ which tends to occur at the ZnO/perovskite interface, {several} approaches have been developed to solve decomposition {issue}: (i) high temperature annealing of ZnO layer\cite{Siempelkamp2015}, (ii) employing polymer interlayers between ZnO/perovskite to avoid the direct interaction\cite{Kim2014,Qiu2015}, and (iii) the use of aluminum (Al) as a dopant in ZnO to passivate the OH$^-$ group\cite{Shen2016,Tseng2016}. {Previously}, Tseng et al. have achieved a PCE of 17.6\%, which was considered to be the {highest} for ZnO-based perovskite solar cell by using {the} sputtered Al-doped ZnO (AZO) thin film as electron transport layer (ETL)\cite{Tseng2016}. {In} the above research, it is worth noting that the use of AZO as ETL could synergistically produce higher efficiency and stability for perovskite solar cells. However, to fabricate better thin film quality of AZO ETLs with high conductivity and crystallinity, preparation by physical vapor deposition (PVD) system with the use of vacuum chamber is required which could lead to an increase in production cost and hinder the scale-up technologies in roll-to-roll fabrication of large area solar cells. Therefore, the development of solution-processed AZO ETLs by sol-gel technique is {desired} for it presents several advantages{, including} facile synthesis, solution-based {growth}, long-term stability, and low temperature processability that are suitable for flexible photovoltaics.

Inspired by these concepts based on effective electron extraction offered by ZnO ETL\cite{Kelly2014} and great passivation ability of AZO\cite{Tseng2016}, we {adopted} a synergistic strategy of {using low-temperature} solution-processed ZnO/AZO bilayer as an ETL which can offer better electron extraction and transportation due to the stair-case band alignment in the ZnO/AZO/perovskite interface. In addition, the ZnO/AZO bilayer configuration {can improve} the interfacial contact to active layer of perovskite {and} suppress the charge recombination, ultimately resulting in enhancement in PCE, $J_{sc}$, and $V_{oc}$ of the device. Detailed investigations on the prominent effect of ZnO/AZO bilayer as an ETL for the inverted PHJ perovskite solar cells were presented. The $V_{oc}$ and $J_{sc}$ values of the devices containing ZnO/AZO bilayer were enhanced compared to those of devices containing only ZnO. Moreover, the transport properties of the ZnO and interfacial AZO layer were systematically investigated by using ultraviolet photoelectron spectroscopy (UPS) and absorption measurements. {Better charge transfer was found between ZnO/AZO bilayer and perovskite which {leads to} higher $V_{oc}$ and $J_{sc}$ in the device.} A maximum PCE of 16.07\% was obtained for inverted planar perovskite solar cells comprising ZnO/AZO bilayer ETL that is ~24\% higher than those containing only ZnO ETL.

\section{Experimental}

\subsection{Syntheses of materials}

Methylammonium iodide (MAI) preparation aqueous HI (57 wt\% in water), methylamine (CH$_3$NH$_2$, 40 wt\% in aqueous solution), PbI$_2$ (99.998\%), dimethyl sulfoxide (DMSO), and diethyl ether were purchased from Alfa Aesar and used without further purification. CH$_3$NH$_3$I was synthesized by the method described in literature\cite{Moorthy2016}. Briefly, CH$_3$NH$_3$I was synthesized by reacting aqueous HI with CH$_3$NH$_2$ at 0$^{\circ}$C for 2 h in three-neck flask under a N$_2$ atmosphere with constant stirring. A white precipitate (CH$_3$NH$_3$I) formed during rotary evaporation of the solvent. The precipitated white powder was collected, washed three times with diethyl ether, and then dried under vacuum at 60$^{\circ}$C overnight. This dried CH$_3$NH$_3$I white powder was eventually stored in a glove box for further fabrication. For the ZnO sol-gel solution which was synthesized according to a modified procedure reported in literatures\cite{Sun2011}. Typically, zinc acetate dihydrate, aluminum chloride hexahydrate, 2-methoxethanol, and monoethanolamine (MEA) were used as the starting materials, solvent, and stabilizer, respectively. Zinc acetate {dihydrate} were first dissolved in a mixture of 2-methoxethanol. The molar ratio of MEA to zinc acetate {dihydrate} was maintained at 1.0 and the concentration of zinc acetate was 0.5 M. In order to prepare Al-doped ZnO sol-gel solutions, the concentration of Al as a dopant were varied at different concentration (1, 3 and 5 \%) with respect to Zn and the concentration of zinc acetate was 0.25 M. When the mixture was stirred, MEA was added drop by drop. Then, the resulting mixture was vigorously stirred in {a cone-shaped} glass beaker and settled in the water bath at 80$^{\circ}$C for three hours. Finally, {the} sol-gel solution was aged for one day to obtain {a} clear and transparent homogeneous solution.

\subsection{Device fabrication and characterization}

Substrates of the cells are fluorine-doped tin oxide (FTO) conducting glass (Ruilong; thickness 2.2 mm, sheet resistance 14 $\Omega$/square). Before use, {the} FTO glass was first washed with mild detergent, rinsed with distilled water {several} times and subsequently with ethanol in an ultrasonic bath, finally dried under air stream. The ZnO/AZO bilayer thin films were sequentially deposited on the FTO glass {substrate via} spin coating method. The coating solution was dropped onto the FTO {substrate} which was rotated at 3000 rpm for 30s, followed by thermal annealing at 200$^{\circ}$C for 30 min on a hot plate to evaporate the solvent and remove organic residuals. Next, the substrates were transferred to a glove box for further deposition of the perovskite active layer through the two-step spin-coating method. PbI$_2$ (48 wt\%)+KCl (1 wt\%) additive were dissolved in 1ml dimethyl sulfoxide; CH$_3$NH$_3$I (4.25 wt\%) was dissolved in 1 ml 2-propanol solvent. Both solutions were kept on a hot plate at 70$^{\circ}$C overnight. A hot PbI$_2$ solution was spin-coated onto ZnO thin film and annealed directly (70$^{\circ}$C, 10 min). The hot CH$_3$NH$_3$I solution was then spin-coated onto the PbI$_2$ film; the structure was kept on the hot plate at 100$^{\circ}$C for 5 min to form a crystalline perovskite film. The hole-transporting spiro-OMeTAD material (5 wt\%), 28.5 $\mu$l 4-tert-butylpyridine (tBP) solution, 17.5 $\mu$l lithium bis(trifluoromethyl-sul-phonyl)imide solution (520 mg in 1 ml acetonitrile) all dissolved in 1 ml chlorobenzene (CB) was further spin-coated on {top}. Finally, the device was completed through sequential thermal evaporation of MoO$_3$ (8 nm) and a silver electrode (80 nm) through a shadow mask under vacuum (pressure: 1$\times$10$^{-6}$ torr). The active area of each device was 10 mm$^2$. Ultraviolet photoelectron spectroscopy (UPS) was performed {by} using a PHI 5000 Versa Probe apparatus equipped with an Al K$\alpha$ X-ray source (1486.6 eV) and He (I) (21.22 eV) as monochromatic light sources. Absorption spectra of the films were measured {with the use of} a Jacobs V-670 UV-Vis spectrophotometer. The morphology of ZnO, ZnO/AZO bilayer and perovskite thin films were investigated by field-emission scanning electron microscope (FEI Nova 200, 10 kV). Crytallographic information was obtained {by} using X-ray diffraction (XRD) on a Bruker D8 X-ray diffractometer (2$\theta$ range: 10 -- 60$^{\circ}$; step size: 0.008$^{\circ}$) equipped with a diffracted beam monochromator set for Cu K$\alpha$ radiation ($\lambda$= 1.54056 \AA). The photocurrent density-voltage (J-V) characteristics of the cells were illuminated inside a glove box by a Xe lamp as a solar simulator (Thermal Oriel 1000 W), which provided a simulated AM 1.5 spectrum (100 mWcm$^{-2}$). The light intensity was calibrated {by} using a mono-silicon photodiode with a KG-5 color filter (Hamamatsu). External quantum efficiency (EQE) spectra were measured under monochromatic illumination (Enlitech, QE-R3011). Devices were encapsulated before they were removed for EQE measurement. The micro-photoluminescence spectra ($\mu$-PL, Horiba Jobin Yvon HR-800) of the ZnO, ZnO/AZO bilayer and perovskite thin films were obtained {by} using a 325 nm He-Cd CW laser as the excitation source with a 2400 grooves/mm grating in the backscattering geometry. Time-resolved PL measurements were carried out by a time-correlated single photon counting (TCSPC) system and samples were photoexcited by using a 405 nm pulse laser source. All of the measurements were carried out at room temperature (RT).

\section{Results and discussion}

Figure 1(a) illustrates the device architecture of PHJ perovskite solar cells with ZnO/AZO bilayer used in this study. The device structure is FTO/ZnO/AZO/CH$_3$NH$_3$PbI$_3$/Spiro-OMeTAD/MoO$_3$/Ag, where the ZnO/AZO bilayer was deposited onto the FTO electrode as ETL. To efficiently extract electrons from CH$_3$NH$_3$PbI$_3$  active layer, a 10-nm thick sol-gel processed AZO layer was grown at the ZnO/perovskite interface. Since AZO has a higher conduction band compared to that of ZnO, with high electron density and faster electron mobility, the ZnO/perovskite recombination would be restrained. The corresponding cross sectional image of FTO/ZnO/AZO/perovskite device was shown in Fig. 1(b).  The ZnO/AZO thin layer was deposited on the FTO substrate by spin-coating method. Then, the CH$_3$NH$_3$PbI$_3$ or perovskite active layer was deposited onto the ZnO/AZO bilayer using two-step spin-coating method. A uniform distribution of perovskite film was formed on the top of the ZnO/AZO bilayer. The PHJ perovskite devices were finally fabricated using spiro-OMeTAD as a hole transporting layer (HTL), MoO$_3$ and silver as top electrode.  {The thicknesses of ZnO and AZO layers were varied to optimize the peformance of solar cells {(See supporting information, Figure S1)}. The best thicknesses of ZnO and AZO layers found are $\sim$30 nm and $\sim$10 nm, respectively.}

\begin{figure}[h]
\centering
  \includegraphics[height=8cm]{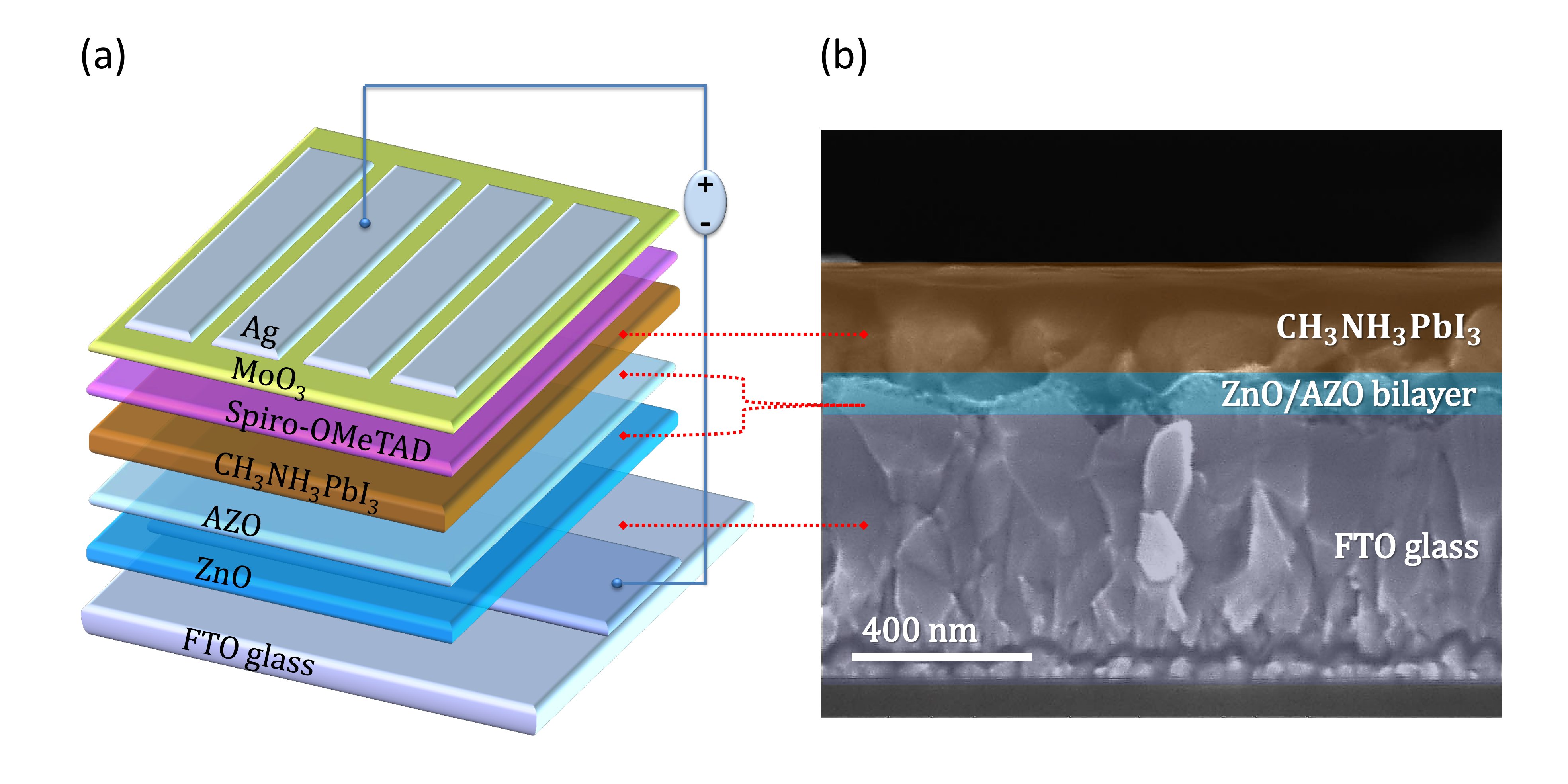}
  \caption{(a) Device architecture of the FTO/ZnO/AZO/CH$_3$NH$_3$PbI$_3$/
Spiro-OMeTAD/MoO$_3$/Ag used in the study.  (b) The corresponding cross-sectional SEM image of FTO/ZnO/AZO/perovskite device.}
  \label{Fig1}
\end{figure}

\begin{figure}[h]
\centering
  \includegraphics[height=11cm]{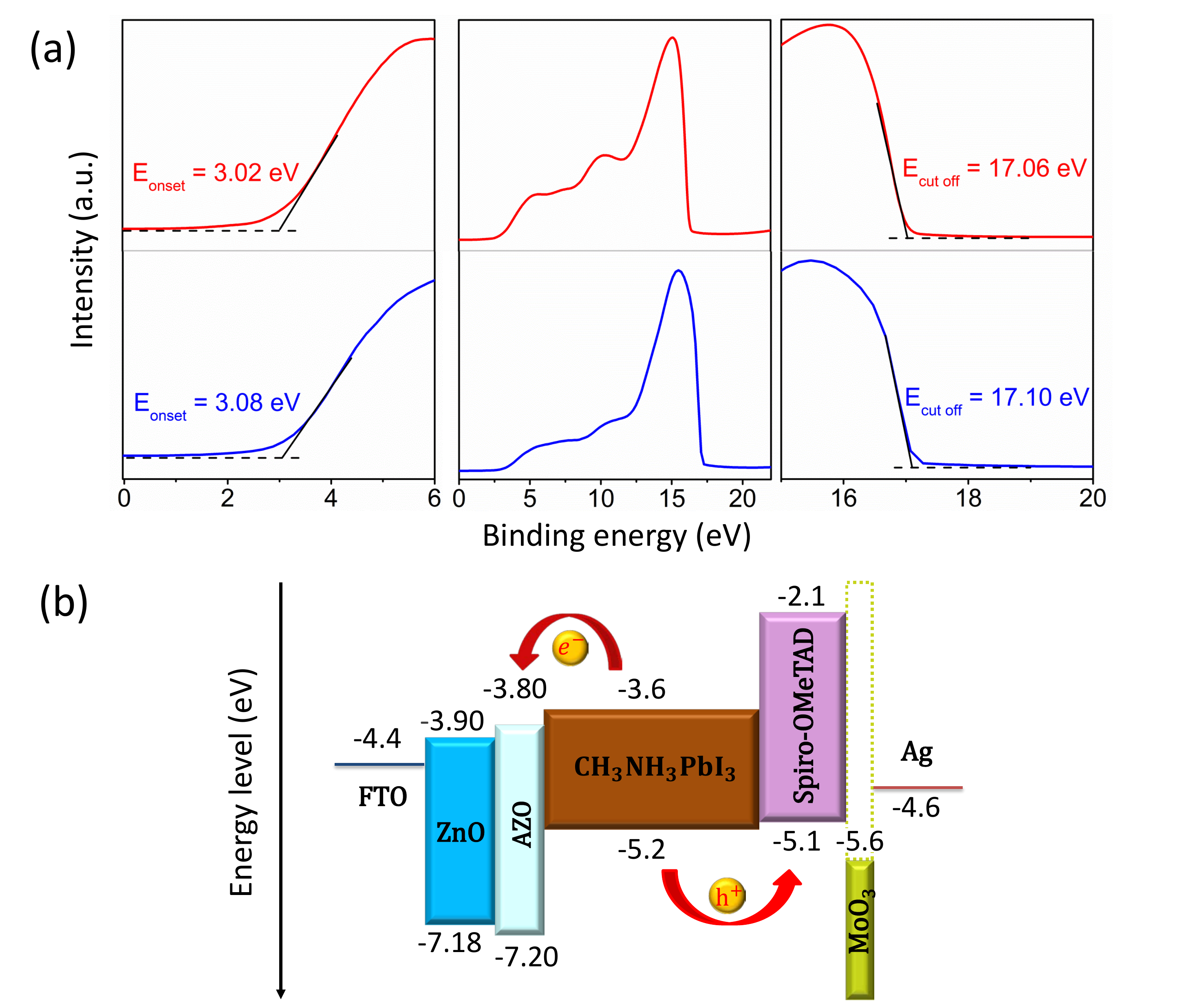}
  \caption{(a) UPS spectra for ZnO layer (red curve) and ZnO/AZO bilayer (blue curve). {Left: onset region, Middle: the whole UPS photoemission spectra. Right: cut off region.} (b) Energy-level diagram of the ZnO/AZO bilayer-perovskite device.}
  \label{Fig2}
\end{figure}

To investigate the electronic structures of ZnO/AZO bilayer and evaluate the effect of the AZO interfacial modification,
UPS and absorption measurements were performed to the ZnO and AZO thin films. Energy-level diagrams of other Al-doping concentrations (3\% and 5\%) can be found in Figure S2. Fig. 2(a) presents the UPS spectra of the ZnO and AZO thin films. The low binding energy tail (E$_{onset}$) of UPS spectra for ZnO and AZO samples are shown in Fig. 2(a).  The VBM energy levels (relative to vacuum level) are calculated by using eq. 1\cite{Hu2013}.

\begin{equation}\label{e1}
-VBM = h\nu - (E_{cutoff}-E_{onset})
\end{equation}

\noindent where $h\nu$ = 21.22 eV is the incident photon energy and the high binding energy cut off ($E_{cutoff}$) spectra of ZnO and AZO obtained from Fig. 2 (a) are 17.06 and 17.10 eV, respectively. The corresponding VBM energy levels of ZnO and AZO are -7.18 and -7.20 eV.  In order to determine the positions of conduction band minimum (CBM) of ZnO and AZO, the optical bandgap energies are required. The optical bandgap energies ($E_g$) of ZnO and AZO were estimated using eq. 2\cite{Mane2005}.

\begin{equation}\label{e2}
{(\alpha h\nu)}^n = A (h\nu -E_g)
\end{equation}

\noindent where n = 1/2 for indirect transition, n = 2 for direct transition; $A$ is proportional constant, $\alpha$ is the absorption coefficient, $h$ is Planck's constant, $\nu$ is frequency of incident photon. Here the $E_g$ was determined by using $({\alpha}{h\nu})^2$ = $A(h\nu-E_g)$ since ZnO and AZO are both direct transition in nature. As shown in {Figure S3}, the corresponding optical bandgap of ZnO and AZO are 3.28 and 3.40 eV which were obtained by extrapolating the linear region of the curves near the onset of the absorption edge to the energy axis.  From the positions of VBM and optical bandgap obtained from the absorption spectra, the estimated conduction band minimum (CBM) positions with respect to vacuum level (at 0 eV) are -3.90 and -3.80 eV for ZnO and AZO, respectively. In general, the optical bandgap broadening of AZO can be addressed to the increase in the carrier concentration which blocks the lowest states in the conduction band based on the Burstein-Moss effect\cite{Lin2008}. The corresponding energy level diagram is illustrated in Fig. 2(b). The Fermi level values of each layer were aligned due to thermodynamic equilibrium that arises when they are combined. It was revealed that the conduction band minimum (CBM) of AZO is slightly higher than that of ZnO ($\Delta$E$_V$ = 0.1 eV), thus indicating that this ZnO/AZO bilayer band alignment could benefit electron extraction from CH$_3$NH$_3$PbI$_3$ active layer, suppress charge accumulation in the ZnO layer, and restrain the charge recombination at ZnO/perovskite interface. These results suggest that the sequencing of ZnO/AZO bilayer structure is influential in tuning the energy level alignment and increase in voltage output of the device.

\begin{figure}[h]
\centering
  \includegraphics[height=8cm]{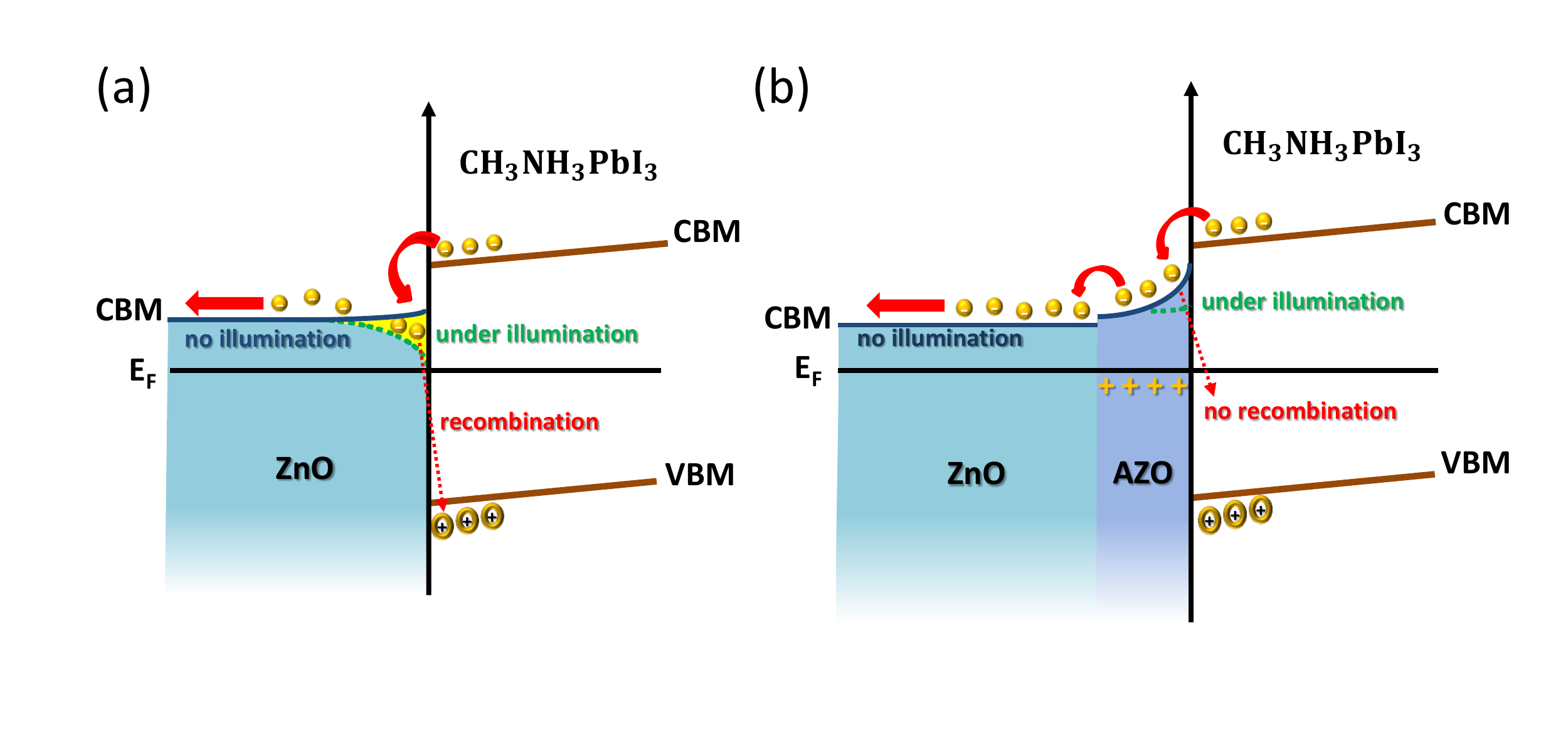}
  \caption{{Schematic diagrams of field-effect-induced band bending in the heterojunctions of (a) ZnO/CH$_3$NH$_3$PbI$_3$ and (b) ZnO/AZO/ CH$_3$NH$_3$PbI$_3$ without illumination (solid blue line) and with illumination (dashed green line). "++++"  in AZO region indicates the positive charges left after the depletion of free electrons.}}
  \label{Fig3}
\end{figure}

{The comparison of band profiles (including the band-bending effect due to doping and charge transfer) of  ZnO/CH$_3$NH$_3$PbI$_3$  and ZnO/AZO/CH$_3$NH$_3$PbI$_3$ interfaces with (green dashed lines) and without (blue solid lines) the effect of electron-hole interaction after photo-excitation are shown in Fig. 3. Since the Fermi level (with respect to vacuum) in  FTO ($n$-contact) is 0.2 eV higher than that of Ag ($p$-contact), electrons must be transferred from the $n$ side to $p$ side in order to make their Fermi levels aligned when FTO and Ag are brought into contact in the solar cell device, leaving surface charges near the ohmic contacts at both ends. This leads to  a nearly constant electric field in the active region. Most of the voltage drop should appear in the perovskite region, since it is intrinsic. The qualitative behavior of band bending depicted in Fig. 3 has taken into account the built-in voltage and the solution to the Poisson equation.

\begin{equation} \partial^2 V(z)/\partial z^2=4\pi e^2 [N_d(z)-n(z)]/\varepsilon(z) \label{Po} \end{equation}

where $V(z), n(z), N_d(z)$, and $/\varepsilon(z)$ denote the  electrostatic potential energy, electron carrier density, dopant concentration, and dielectric constant at position $z$, respectively. In Fig. 3(a), the band profile of ZnO has a weak band bending, since the built-in electric field is partially screened by the free carriers in ZnO, which is slightly n-type.  An extra electron trapping potential can be created (as indicated by the dotted line) by the electron-hole attraction after photo-excitation. These trapped electrons in the triangular well can hinder the charge separation process and part of them will recombine with holes in CH$_3$NH$_3$PbI$_3$. In Fig. 3(b), a thin AZO layer (which has higher free electron density  than ZnO due to Al doping) is inserted between ZnO and CH$_3$NH$_3$PbI$_3$. Since the conduction band minimum (CBM) of AZO  is higher than the CBM of ZnO, free electrons must be transferred from AZO layer to the ZnO/FTO interface region, leaving a depleted AZO region with net positive charges of dopants, which are assumed to be uniformly distributed.  Consequently, there is a strong band bending in the AZO region after adding the self-consistent potential which satisfies the Poisson equation [Eq.~(\ref{Po})], as seen in Fig. 3(b). The curvature of the band bending is proportional to the dopant concentration due to Eq.~(\ref{Po}). After photo-excitation, the attractive electron-hole interaction will reduce the band bending, but not enough to form a trapping potential at the AZO/perovskite interface. Thus, the charge separation process in ZnO/AZO/perovskite interface can be enhanced in comparison to the ZnO/perovskite interface.}

\begin{figure}[h]
\centering
  \includegraphics[height=10cm]{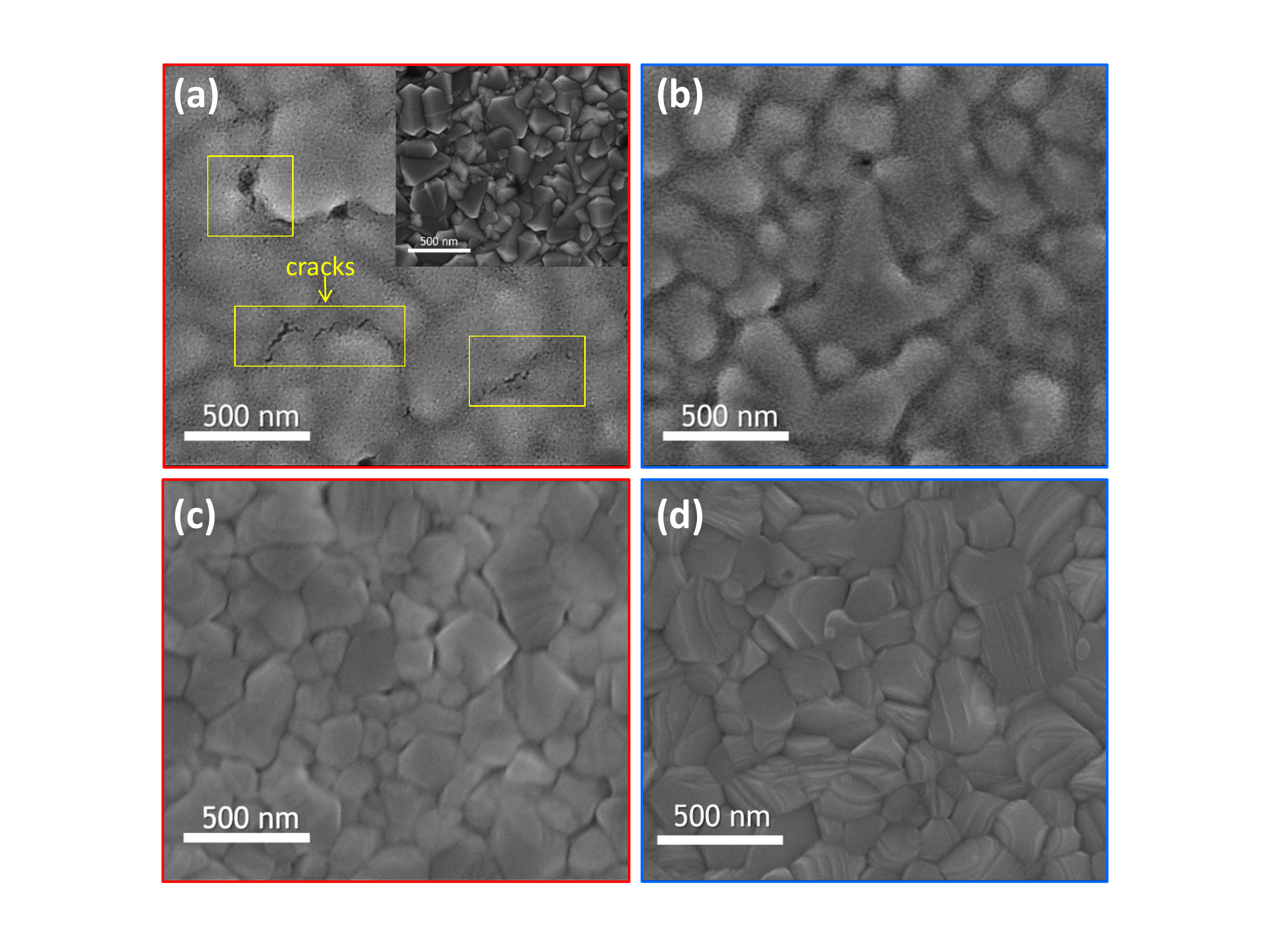}
  \caption{SEM top-view images of (a) ZnO (inset is the bare FTO substrate); the areas of visible cracks were {indicated by} yellow boxes, (b) ZnO/AZO, (c) perovskite film grown on the ZnO and (d) perovskite film grown on ZnO/AZO.}
  \label{Fig4}
\end{figure}

The surface morphologies of the {pure ZnO ETL, the ZnO/AZO bilayer ETL}, and the active layer of CH$_3$NH$_3$PbI$_3$, respectively, are shown in Fig. 4. As observed in Fig. 4(a), the morphology of ZnO monolayer on {FTO} substrate shows a clear sol-gel processed  {granular} feature. However, several cracks {(indicated by yellow boxes)} can be found in the {ZnO layer which correlate with the FTO grain boundaries. These cracks} could lead to direct contact {between  FTO and} perovskite, causing serious charge recombination.  {On the other hand, with the AZO thin film modification} the ZnO/AZO bilayer [Fig. 4(b)] {forms a} smooth and homogenous ETL surface without visible cracks. Figure 4(c-d) show {top-view} SEM images of perovskite films deposited on ZnO and ZnO/AZO bilayer thin films. The less-compact perovskite morphology observed in the film  grown on ZnO {[Fig. 4(c)]} may result in poorer connectivity between adjacent crystallites, leading to a more tortuous pathway for carrier transport and a greater likelihood of charge recombination\cite{Liu2014}. In contrast, the uniform and {densely}-packed perovskite film [Fig. 4(d)] with grain sizes at approximately $\sim$ 150--450 nm was formed on {top of the} ZnO/AZO bilayer film.

\begin{figure}[h]
\centering
  \includegraphics[height=9cm]{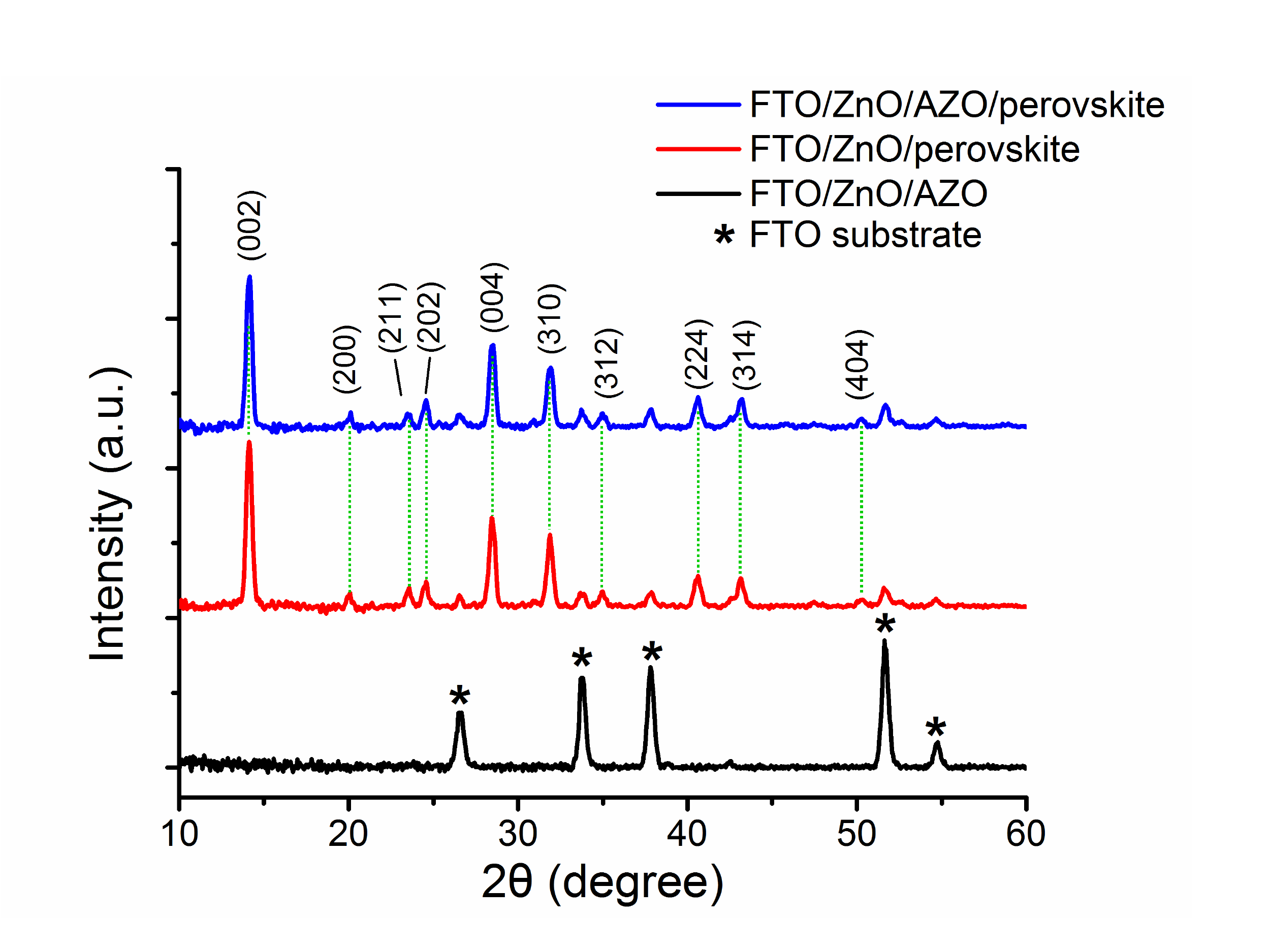}
  \caption{XRD patterns of FTO/ZnO/AZO/perovskite, FTO/ZnO/perovskite, and FTO/ZnO/AZO.}
  \label{Fig5}
\end{figure}

The XRD patterns of {FTO/ZnO/AZO/perovskite, FTO/ZnO/perovskite, and FTO/ZnO/
AZO} are displayed in Fig. 5. The sol-gel-processed ZnO/AZO bilayer film shows an amorphous structure with no crystallinity (2$\theta$ in the range of 10 to 60 degrees) which is consistent with {results in} the literature\cite{Sun2011} {for} low-temperature-processed ZnO ETLs. The diffraction pattern of CH$_3$NH$_3$PbI$_3$ film reveals peaks {associated with} the (002), (211), (202), (004), (310), (312), (224), (314) and (404) planes, respectively, confirming the formation of a tetrahedral perovskite crystal structure. The peak positions indicate that a pure perovskite phase was obtained; no secondary phases {arising} from incomplete formation of perovskite {appeared} (e.g., PbI$_2$ or CH$_3$NH$_3$I).

\begin{figure*}[h]
\centering
  \includegraphics[height=11cm]{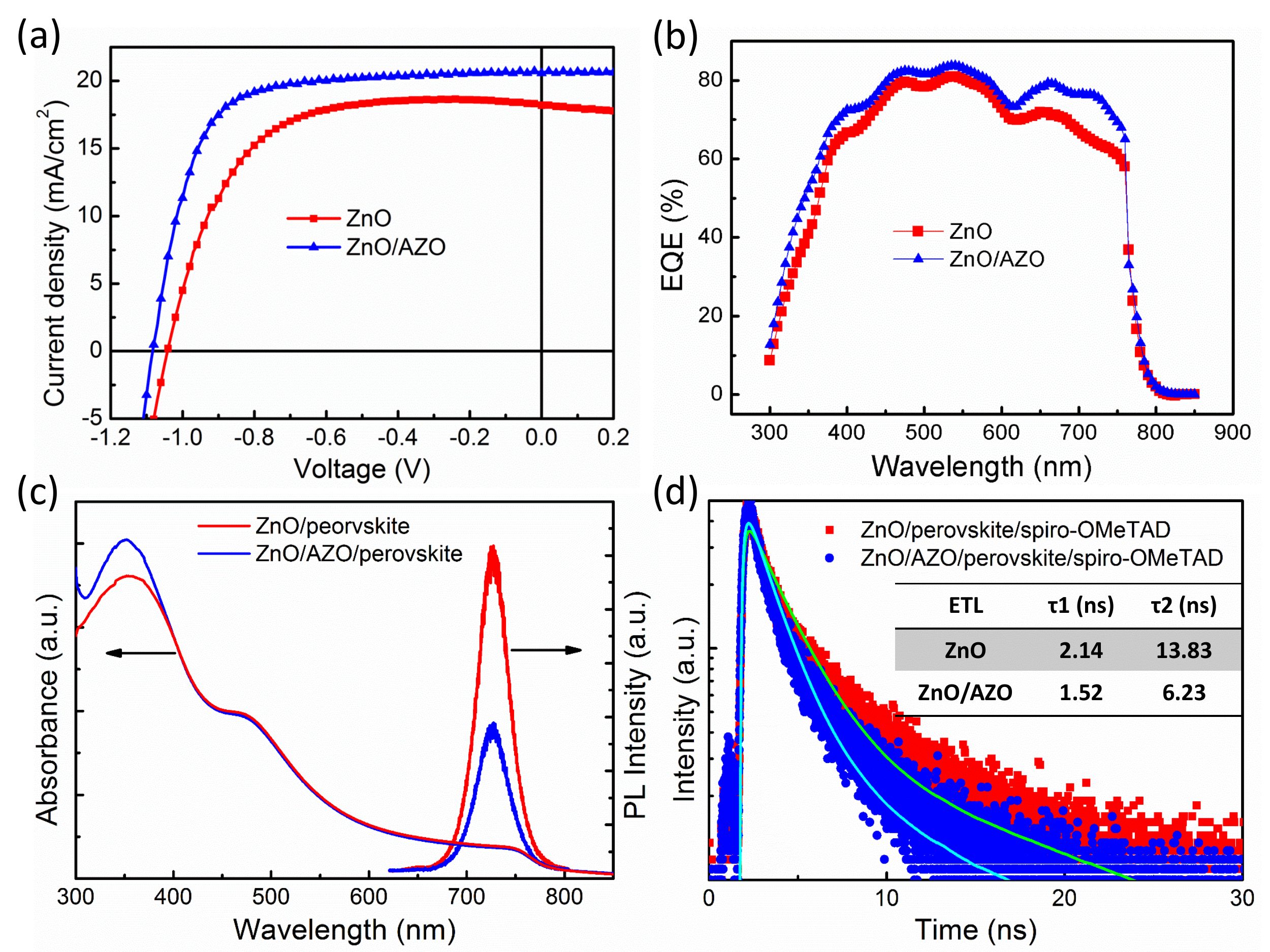}
  \caption{(a) J-V characteristics of perovskite solar cells incorporating ZnO and ZnO/AZO as ETLs measured under AM 1.5 G illumination (100 mW cm$^{-2}$. (b) EQE spectra of the corresponding cells. (c) Absorption and steady-state PL spectra of perovskite films. (d) Time-resolved PL spectra of perovskite films comprising with ZnO and ZnO/AZO ETLs.}
  \label{Fig6}
\end{figure*}

\begin{table*}
  \caption{Photovoltaic performance parameters of  the {best performing} devices  prepared with {only ZnO layer} and {with} ZnO/AZO bilayer as ETLs}
  \label{tbl:photovoltaic}
 \begin{tabular}{c c c c c c c}
    \hline
 \bf{Cathode}
& \multicolumn{1}{p{1.75cm}} {\centering $V_{oc}$ \\ ($V$)}
& \multicolumn{1}{p{1.75cm}}{\centering $J_{sc}^a$ \\ ($mA{\cdot}cm^{-2}$)}
& \multicolumn{1}{p{1.75cm}}{\centering $J_{sc}^b$ \\ ($mA{\cdot}cm^{-2}$)}
& \multicolumn{1}{p{1.75cm}}{\centering FF \\ (\%)}
& \multicolumn{1}{p{1.75cm}}{\centering PCE \\ (\%) }
& \multicolumn{1}{p{1.75cm}}{\centering $R_{SH}$ \\ ($\Omega{\cdot}cm^2$)} \\
    \hline
    \bf{FTO/ZnO} & {1.04} & 18.20 & 17.62 & 64.7 & 12.25 & 394 \\
    \bf{FTO/ZnO/AZO} & 1.09 & 20.58 & 19.55 & 71.6 & 16.07 & 21153\\
    \hline
  \end{tabular}

$^{a}$ Measured $J_{sc}$ values from the AM 1.5 G solar simulator.
$^{b}$ Integrated $J_{sc}$ values from the EQE measurement.
$^{c}$ The histograms {for photovoltaic parameters of} 50 cells are illustrated in Fig. S5.

\end{table*}

The J-V characteristics and EQE (External Quantum Efficiency) spectra of the cells based on ZnO and ZnO/AZO bilayer ETLs are shown in Figs. 6(a) and 6(b). The photovoltaic parameters of devices {with best performance} are summarized in Table 1. The device employing ZnO/AZO bilayer ETLs shows significant enhancement in photovoltaic parameters with respect to that of ZnO monolayer ETL. The comparison of AZO ETLs {with different Al-doping concentrations (1, 3, 5\%) is shown in Fig. S4.} The device with ZnO/AZO bilayer presents distinct {improvement} in $J_{sc}$ (from 18.20 to 20.58 $mA{\cdot}cm^{-2}$) and $V_{oc}$ (from 1.04 to 1.09 V), resulting in the best PCE (from 12.25 to 16.07\%). Notably, the {fill-factor (FF)} value was {also} largely improved from 64.7 to 71.6\% by adding AZO as {the} buffer layer. Based on the slope $(dJ/dV)^{-1}$ of the corresponding J-V curves at $J_{sc}$, {we obtained a} low shunt resistance ($R_{SH}$) of 394 $\Omega{\cdot}cm^2$ {for} the ZnO monolayer film {and a} high $R_{SH}$ of 21153 $\Omega{\cdot}cm^2$ {for} the ZnO/AZO film. {This translates to a} significant improvement in FF. The increased $J_{sc}$ in ZnO/AZO bilayer is supported by the enhancement in EQE {[see Fig. 6(b)]}. It is found that the PHJ perovskite device comprising of ZnO/AZO bilayer ETLs {exhibits} a higher spectral response with a maxima EQE of 83.7\% in contrast to a lower EQE of 80.9\% for the device comprising of ZnO ETL. {Moreover}, the increased EQE of the device based on ZnO/AZO bilayer ETLs is due to the higher light absorption {within} the range from 300 to 400 nm [See Fig. 6(c)] and the better charge collection efficiency. The discrepancy ($\sim{5}\%$) between the integrated  $J_{sc}$ from EQE and  $J_{sc}$ obtained from the J-V curve is generally observed in the perovskite cell due to the {difference in} measurement methods. In order to investigate the origin of enhanced $J_{sc}$, {PL} measurements were carried out to study the capability for the electron {transfer} from perovskite to ETLs. As shown in Fig. 6(c), the strong PL quenching occurred in perovskite thin film grown on ZnO/AZO bilayer ETL compared to that grown on ZnO ETL indicates high electron transfer efficiency and exciton dissociation of the ZnO/AZO bilayer ETL configuration. Therefore, the higher PL quenching in the perovskite film based on ZnO/AZO bilayer suggests the faster deactivation of the excited state by the efficient charge transfer between perovskite and ZnO/AZO bilayer and thus giving rise to enhanced J$_{sc}$. To further confirm the faster electron transport properties in ZnO/AZO bilayer ETL, the time-resolved PL measurements were conducted with results summarized in Fig. 6(d). The PL lifetime was fitted with a biexponential decay function of the form, $I(t) = A_{1} exp(-t/\tau_{1}) + A_{2} exp(-t/\tau_{2})$, where $I(t)$ is the the intensity at time t after the excitation pulse. By fitting a biexponential function to the fluorescence decay, the lifetime components $\tau_{1}$ and $\tau_{2}$, as well as the amplitudes $A_1$ and $A_2$, can be recovered. The resulting decay lifetimes are listed in the insert of Fig. 6 (d). The fast decay ($\tau_{1}$) process was considered to be the result of the free carriers in the perovskite through transport to ETL or HTL, whereas the slow decay ($\tau_{2}$) process to be the result of radiative decay\cite{Shen2016}. The $\tau_1$ of ZnO/AZO/perovskite/spiro-OMeTAD sample decreased to 1.52 ns, which was much smaller than that of the ZnO/perovskite/spiro-OMeTAD (2.14 ns). This indicated that charger transfer between perovskite and ZnO/AZO bilayer is faster than that without AZO interlayer. The results of PL quenching and lifetime are consistent with the more efficient charge separation due to the stair-case band alignment of our device structure with ZnO/AZO bilayer ETL. It is thus evident that the electrical properties are strongly influenced by employing AZO as the interlayer.

Electrochemical impedance measurements were carried out for perovskite solar cells containing ZnO and ZnO/AZO as ETLs to investigate its charge carrier transport properties systematically. Fig. 7 shows Nyquist plots (imaginary versus real part of the impedance) of perovskite solar cells containing ZnO and ZnO/AZO as ETLs under 1 sun illumination at applied bias voltage of $V=1 V$. The impedance spectra consist of  high frequency (low Z') and low frequency (high Z')  features. In most cases, these features were represented by an arc or semicircle depending on the sample and the applied voltage. In the perovskite solar cell containing only ZnO as ETL, a high resistance represented by a large arc was observed. Higher resistance could be attributed from the presence of cracks [see SEM image in Fig. 3(a)] that allowed direct contact between FTO and perovskite, causing deleterious charge recombination that could yield to poor device performance. In contrast, for the perovskite solar cells containing ZnO/AZO as ETLs, the addition of AZO decreased the resistance {(corresponding to smaller arc in Fig. 7) thus indicating a smoother carrier} movement that leads to less recombination and better device performance.

\begin{figure}[h]
\centering
  \includegraphics[height=8cm]{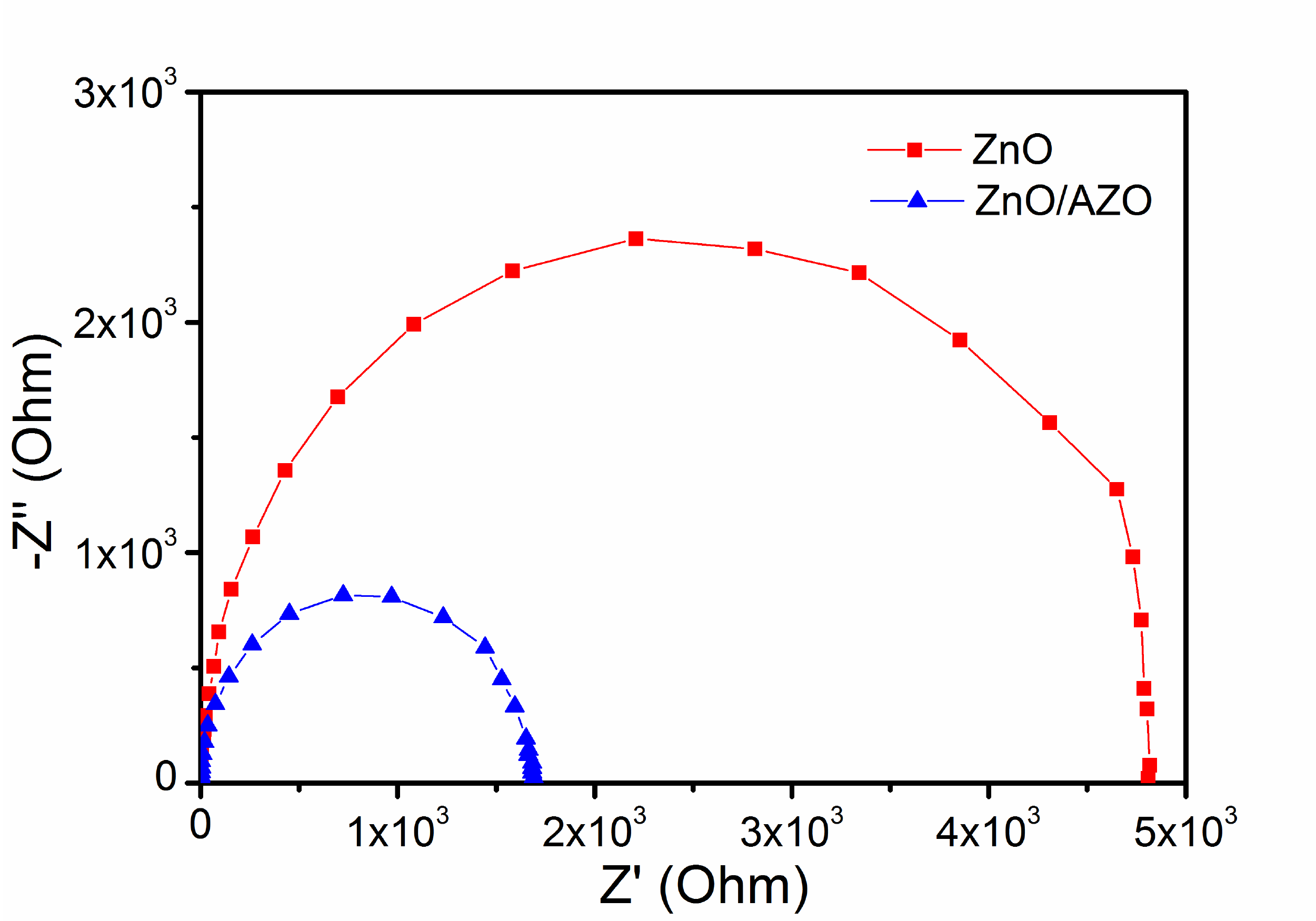}
  \caption{Nyquist plots of perovskite solar cells containing ZnO and ZnO/AZO ETLs under 1 sun illumination with applied bias voltage of 1 V.}
  \label{Fig7}
\end{figure}

\begin{figure}[h]
\centering
  \includegraphics[height=12cm]{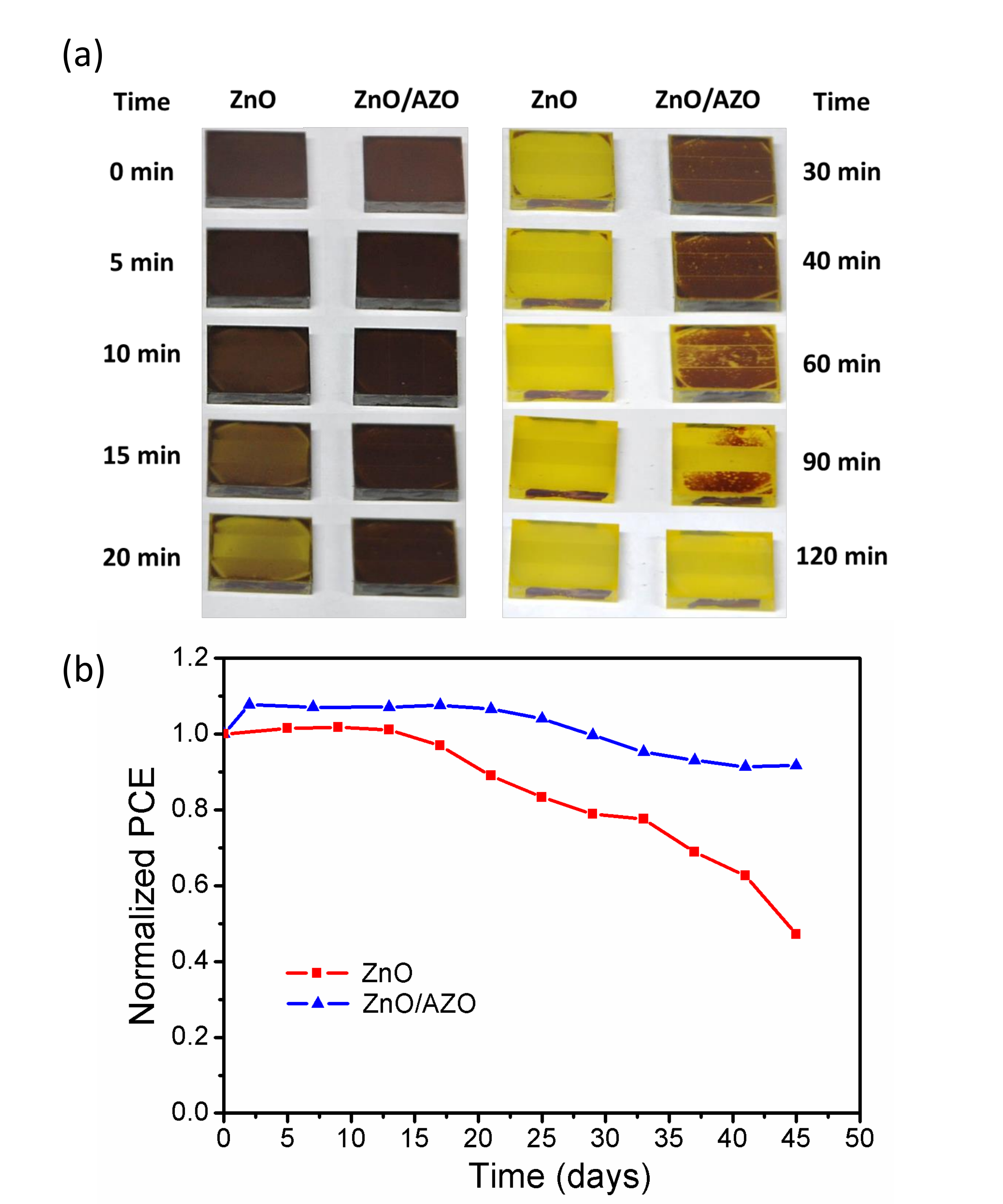}
  \caption{(a) Thermal stability test (with heating at 100$^{\circ}$C) and (b) normalized PCE as a function of time (in days) for perovskite solar cells containing ZnO and ZnO/AZO ETLs.}   \label{Fig8}
\end{figure}

Figure 8 (a) shows the thermal stability test of perovskite films deposited on ZnO and ZnO/AZO ETLs (heating at 100$^{\circ}$C). We found that the CH$_3$NH$_3$PbI$_3$ film deposited on AZO interlayer shows better thermal stability than that deposited on ZnO ETL. The instability of CH$_3$NH$_3$PbI$_3$ film deposited on ZnO ETL was mainly due to the reaction of CH$_3$NH$_3$$^+$ proton in the perovskite layer with the surface hydroxyl groups and residual acetate ligands of sol-gel processed ZnO\cite{Siempelkamp2015}. Previous reports on the thermal stability test of perovskite solar cells with only ZnO ETL confirmed their severe instability against annealing time. Thus, the addition of AZO interlayer improves its thermal stability, which can be ascribed to the decrease in Lewis acid-base chemical reaction between the perovskite and ZnO since the basic property of ZnO has been weakened by doping of aluminum \cite{Shen2016,Tseng2016}. Moreover, we compared the durability of the unencapsulated perovskite solar cells containing {pure ZnO ETLs with those containing ZnO/AZO ETLs} in Fig. 8 (b). {(All testing devices were stored inside a glove box with 30 ppm oxygen and 0.1 ppm water  for 45 days, and their performance was constantly monitored). The device containing ZnO/AZO as ETL exhibits long-term stability with only 9\% degradation of its initial PCE after being kept in the dark for over 45 days. In contrast, the stability of the device prepared with the ZnO as ETL shows deterioration in performance after 12 days of storage with 53\% degradation of its initial PCE. The significant degradation probably arises from the ZnO/perovskite interface (excluding the oxidation effect of spiro-OMeTAD and silver electrodes). To further understand the origin of the decomposition of perovskite layers on different interfaces, XPS measurement was carried out to analyze the $O$ 1s core level spectra for both ZnO and AZO samples (as shown in Figure S6). The lower binding energy of $O_{Zn-O}$ peak (531.08 eV) that is attributed to $O$ atoms in the wurtzite structure of ZnO \cite{Ambade2015} for AZO thin film shifts toward lower energy by 0.11 eV in comparison with ZnO thin film. Moreover, the intensity of $O_{Zn-O}$ peak in AZO thin film is lower than ZnO thin film which can be explained by the Al-doping \cite{Shen2016}. The higher binding energy of second peak (532.97 eV) corresponds to the OH$^-$ or O$_{2}^{-}$ ions in the AZO shifts toward higher binding energy and has lower intensity than in ZnO thin film. The above result indicates that AZO thin film is more oxygen deficient. Thereby the OH$^-$ group passivation enhances the thermal stability and durability of perovskite solar cells containing ZnO/AZO ETL.}

\section{Conclusion}
In summary, we have successfully demonstrated how the low temperature solution processed ZnO/AZO bilayer ETL improves the efficiency of PHJ perovskite solar cell. In this study, the PCE obtained can exceed 16\%. The introduction of AZO layer not only helps the formation of uniform surface morphology to fill the defects but also lead to a stair-case band alignment in the  ZnO/AZO/perovskite heterostruucture, which is shown to yield better electron extraction, thereby significantly enhance the device performance. It is noteworthy that the results obtained for our optimized device surpass the state-of-the-art PCE and V$_{oc}$ of solution processed ZnO-based perovskite solar cells. Moreover, the device prepared with ZnO/AZO bilayer ETLs showed long-term stability; only 9\% degradation of its initial PCE after being kept in the dark for over 45 days. This approach pave a new way for the development of low-cost perovskite solar cells.

\begin{suppinfo}

J-V curves at different molar concentrations of ZnO and ZnO/AZO ETLs, Tauc plots for optical bandgap extraction, UPS spectra of 3\% AZO and 5\% AZO films and corresponding energy-level diagram, J-V curves at different concentrations of AZO interlayers, and Histograms of device performance parameters.

\end{suppinfo}

\section{Author Information}

\subsection{Corresponding Authors}
*Emails: gchu@gate.sinica.edu.tw; yiachang@gate.sinica.edu.tw

\subsection{Author Contributions}
The manuscript was written through contributions of all
authors. All authors have given approval to the final version of
the manuscript.

\subsection{Notes}
The authors declare no competing financial interest.

\begin{acknowledgement}

This work was supported in part by the Ministry of Science and Technology of Taiwan under Contract nos. MOST 104-2112-M-001-009-MY2 and 104-2221-E-001-014-MY3. Additionally, C. W. Chu thanks the Career Development Award of Academia Sinica, Taiwan (103-CDA-M01) for financial support.

\end{acknowledgement}

\bibliography{ref}

\end{document}